\documentclass[11pt]{article}
\usepackage{Blois,epsfig}

\def\Journal#1#2#3#4{{#1} {\bf #2}, #3 (#4)}

\def\NPB{{\em Nucl. Phys.} B}
\def\PLB{{\em Phys. Lett.}  B}

\def\PRD{{\em Phys. Rev.} D}

\def\EPJC{{\em Eur. Phys. J.} C}

\newcommand{\be}{\begin{equation}}
\newcommand{\ee}{\end{equation}}
\newcommand{\ba}{\begin{eqnarray}}
\newcommand{\ea}{\end{eqnarray}}
\newcommand{\ci}[1]{\cite{#1}}

\def\vb0{{\bf b}_0}

\def\gev{\,{\rm GeV}}

\def\xbj{x_{\rm Bj}}
\newcommand{\sla}{\hspace*{-0.20cm}/}

\newcommand{\lsim}{\raisebox{-4pt}{$\,\stackrel{\textstyle
                                                         <}{\sim}\,$}}

\newcommand{\req}[1]{(\ref{#1})}
\def\xb{\bar{x}}

\def\={\,=\,}
\def\eps{\epsilon}

\begin{document}
\vspace*{2cm}
\begin{center}
\Large{\textbf{XIth International Conference on\\ Elastic and
    Diffractive Scattering\\ 
Ch\^{a}teau de Blois, France, May 15 - 20, 2005}}
\end{center}

\vspace*{2cm}
\title{$\phi$ MESON ELECTROPRODUCTION AT SMALL BJORKEN-x}

\author{P.\ Kroll}

\address{Fachbereich Physik, Universit\"at Wuppertal,\\
D-42097 Wuppertal, Germany}

\maketitle\abstracts{
It is reported on an analysis of $\phi$-meson electroproduction at small 
Bjorken-$x$ ($\xbj$) within the handbag approach. The amplitudes factorize 
into generalized parton distributions (GPDs) and a partonic subprocess, 
electroproduction off gluons.  Cross sections and spin density matrix 
elements (SDMEs) are evaluated for $\phi$-meson electroproduction and 
found to be in fair agreement with recent HERA data.}

\begin{figure}[t]
\includegraphics[width=.40\textwidth,bb=112 506 350 715,clip=true]{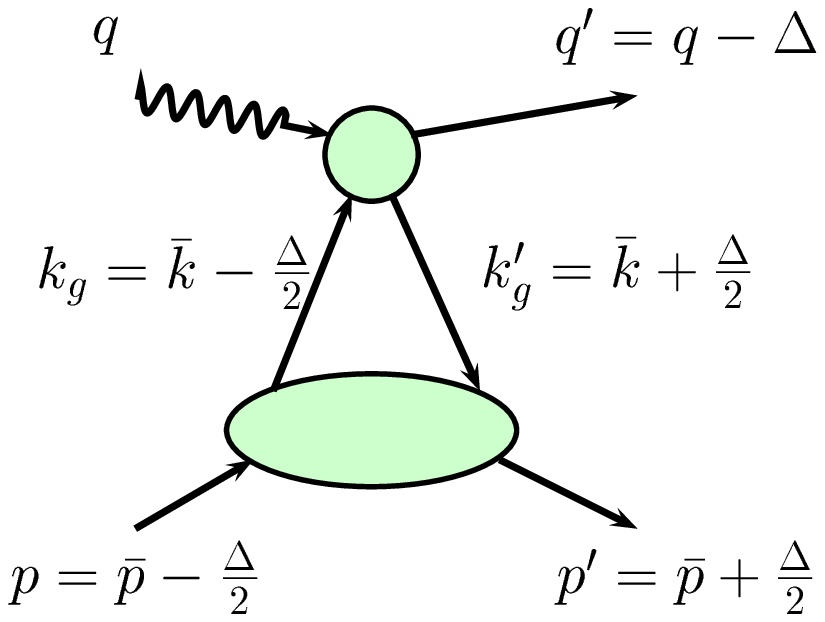}
\hspace*{2.2cm}
\includegraphics[width=.40\textwidth,bb=63 355 530 747,clip=true]{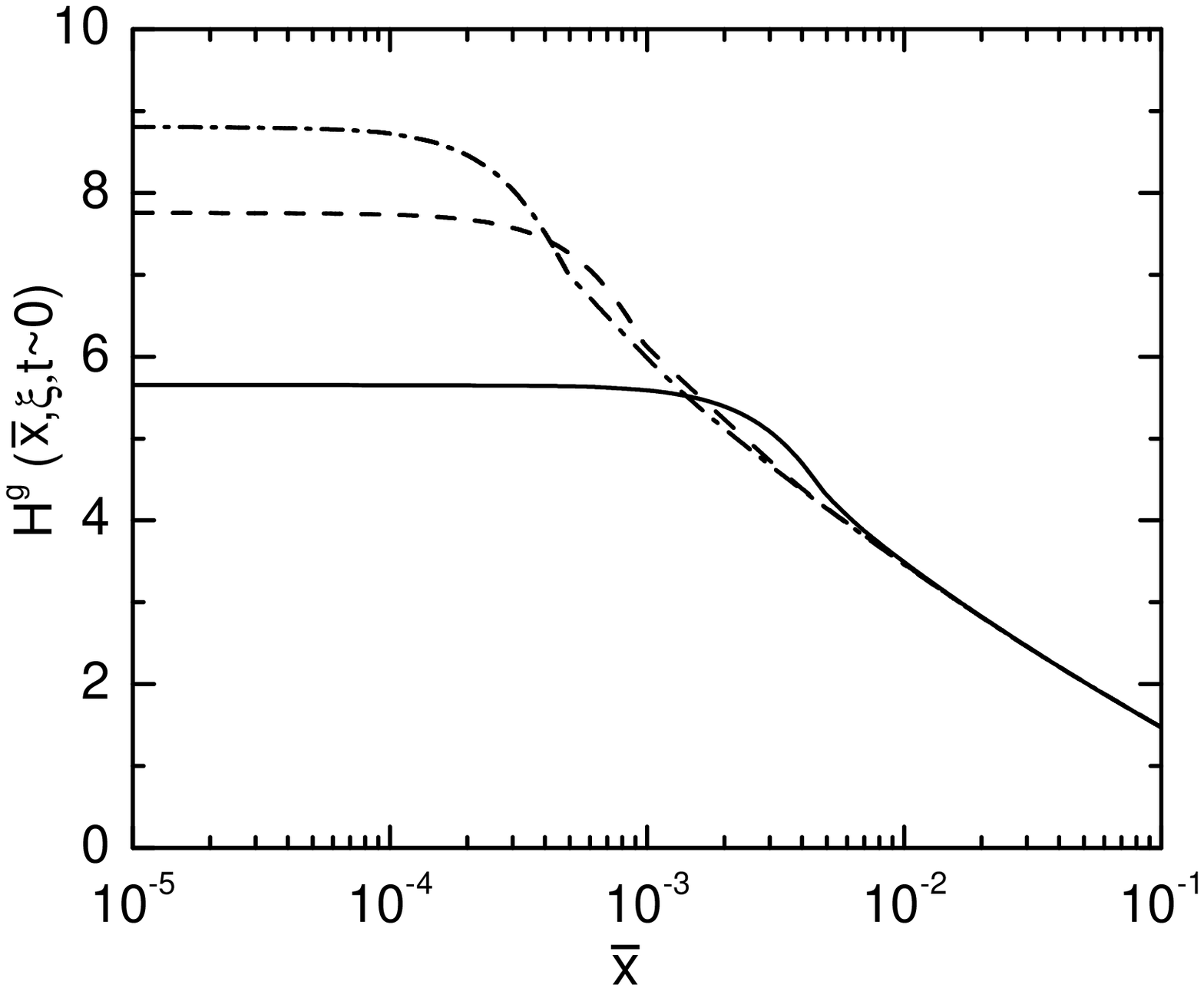}
\caption{Left: The handbag diagram for meson electroproduction off
protons. The large blob represents a GPD while the small one stands
for the subprocess. Particle momenta are specified. Right: The GPD
$H^g$ at $t\simeq 0$. The solid (dashed, dash-dotted) line represents
the GPD at $\xi= 5\;(1\,,\; 0.5)\, \cdot 10^{-3}$ and at a scale of $2\,\gev$.}
\label{fig:1}
\end{figure}

It has been shown~\cite{rad96} that, at large photon  virtuality 
$Q^2$, meson electroproduction factorizes into a partonic subprocess, 
electroproduction off gluons or quarks, $\gamma^* g(q)\to M g(q)$, and
GPDs, representing soft proton matrix elements (see Fig.~\ref{fig:1}). 
At small $\xbj$ and in particular for $\phi$-meson production the quark 
subprocesses can be ignored and only the gluonic subprocess,
$\gamma g \to \phi g$, contributes. In the following I am going to
report on an analysis~\ci{golo} of $\phi$-meson electroproduction
within this handbag approach carried through in the kinematical regime 
of large $Q^2$ and large energy $W$ in the photon-proton c.m.s. but
small $\xbj$ and Mandelstam $t$.\\ 
The structure of the proton is rather complex. In correspondence to
its four form factors there are four gluon GPDs $H^g$, $E^g$,
$\widetilde{H}^g$ and $\widetilde{E}^g$ and four for each quark
flavour. All GPDs are functions of three variables, $t$, skewness
$\xi$ and the average momentum fraction $\xb$, the latter two are 
defined by (see Fig.~\ref{fig:1}) 
\be 
\xi \= \frac{(p-p')^+}{(p + p')^+}\,, \qquad\qquad \xb\= \bar{k}^+/\bar{p}^+\,.
\ee
The skewness is kinematically fixed to $\xi\simeq\xbj/2$ in a small $\xbj$
approximations and in the $\gamma^*p$ c.m.s. This is to be contrasted with the
frequently used leading $\log(1/\xbj)$ approximation \ci{brodsky} where $\xi=0$
and $\xb=\xbj$ is assumed and the GPD replaced by the usual gluon distribution
$g(\xb)$.

\begin{figure}[pt]
\includegraphics[width=.30\textwidth,bb=40 314 533 800,clip=true]
{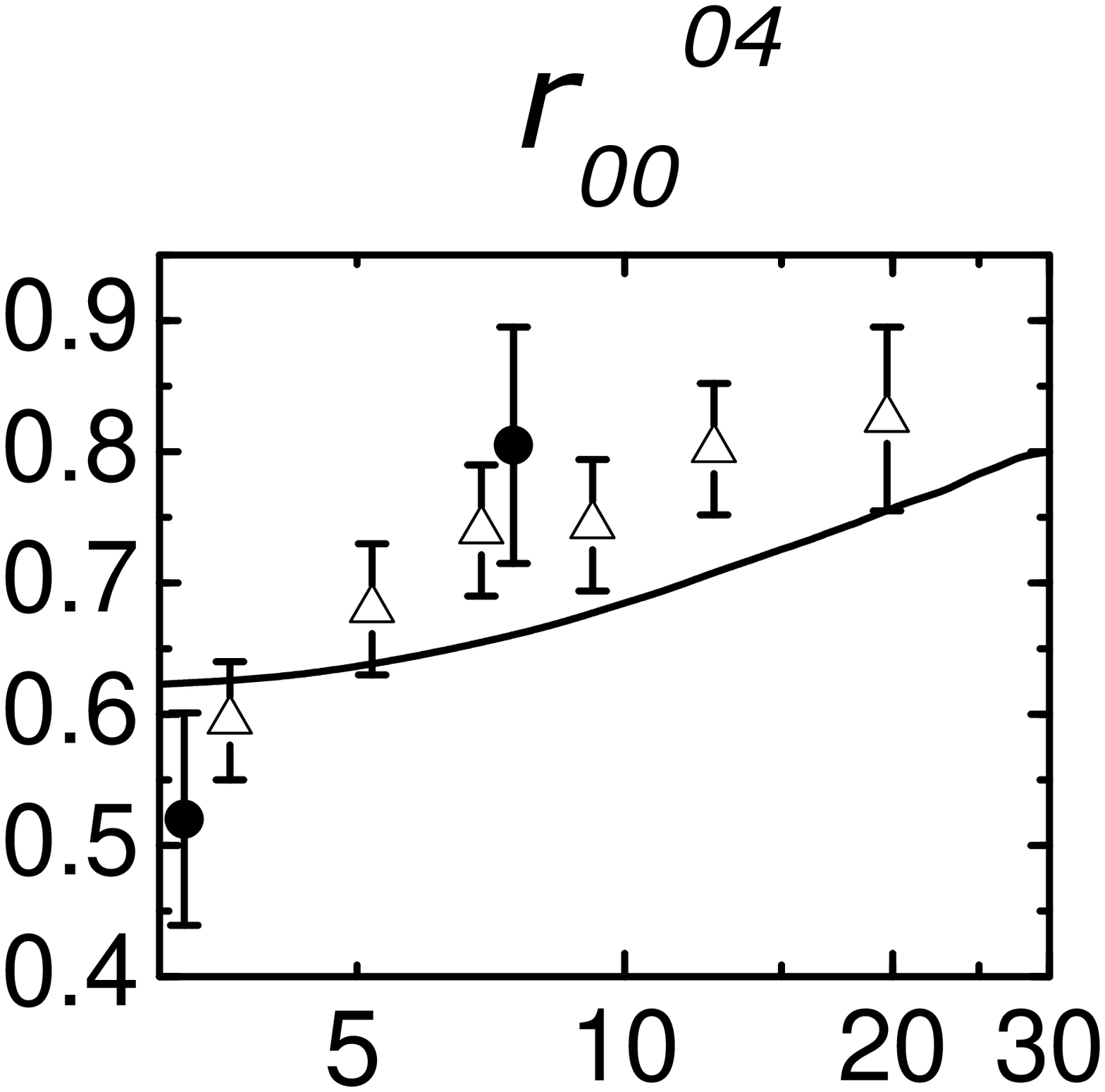}\hspace*{0.05cm}
\includegraphics[width=0.30\textwidth, bb= 62 316 557 800,clip=true]
{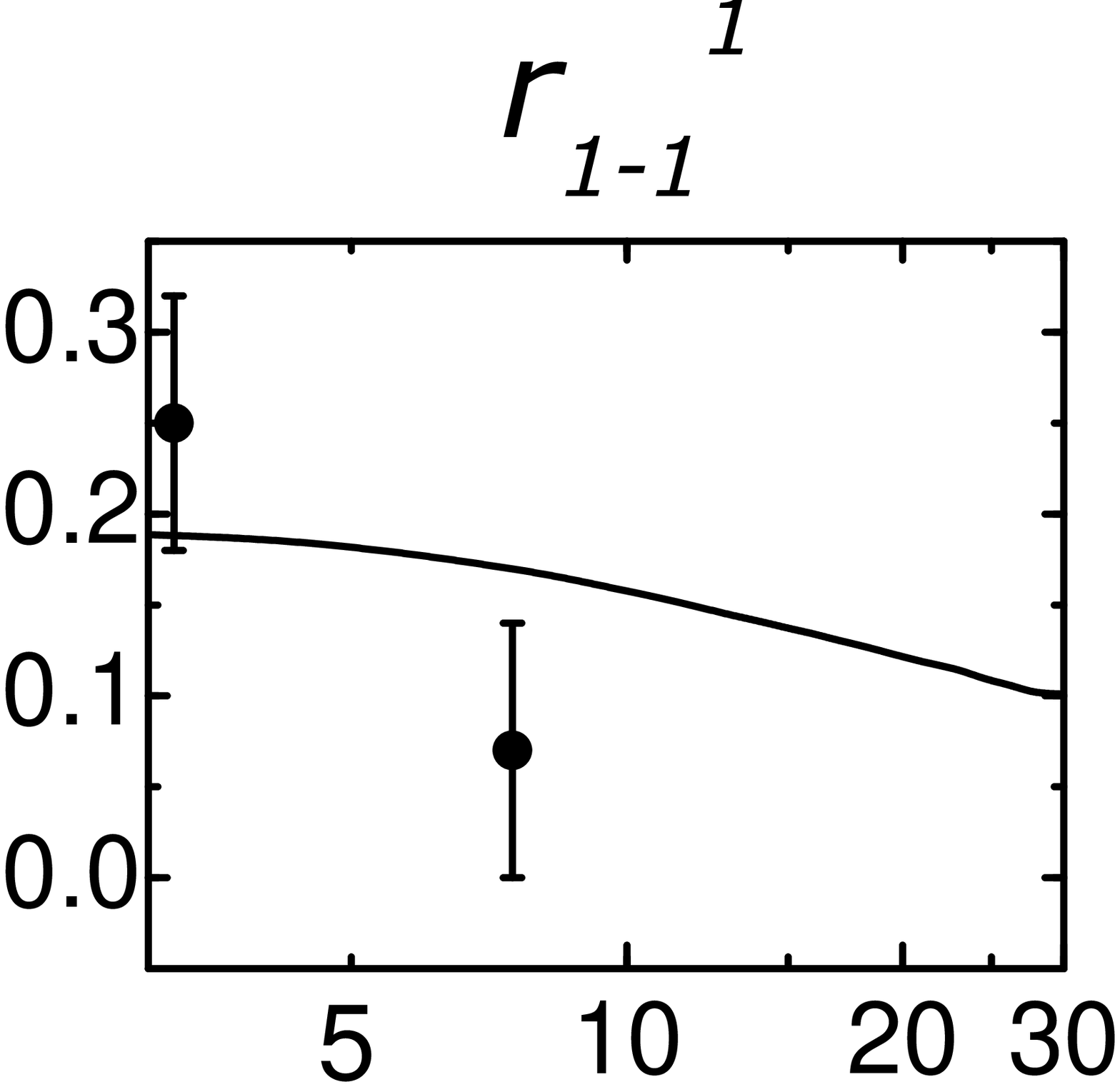}\hspace*{0.05cm}
\includegraphics[width=.30\textwidth,bb=35 322 545 790,clip=true]
{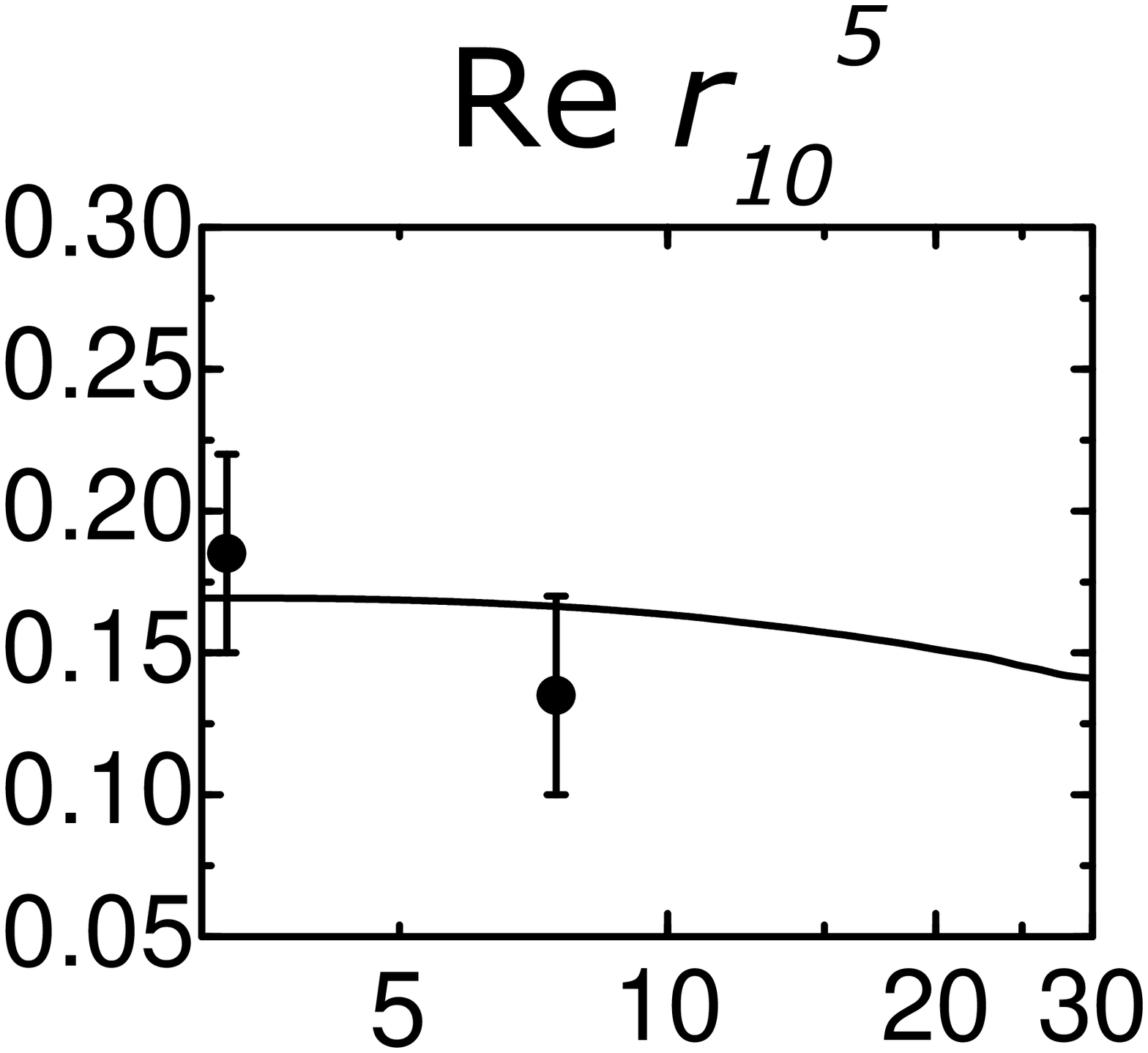}
\caption{Spin density matrix elements of electroproduced $\phi$ mesons
  versus $Q^2$ at $W\simeq 75\,\gev$ and $t\simeq -0.15\,\gev^2$. The
  solid lines are the predictions given in Ref.~\protect\ci{golo}. Data taken
  from Ref.~\protect\ci{h1} ($\bullet$) and \protect\ci{zeus05} ($\triangle$).}
\label{fig:3}
\end{figure}

\begin{figure}[pt]
\includegraphics[width=.44\textwidth,bb=26 332 554 741,clip=true]
{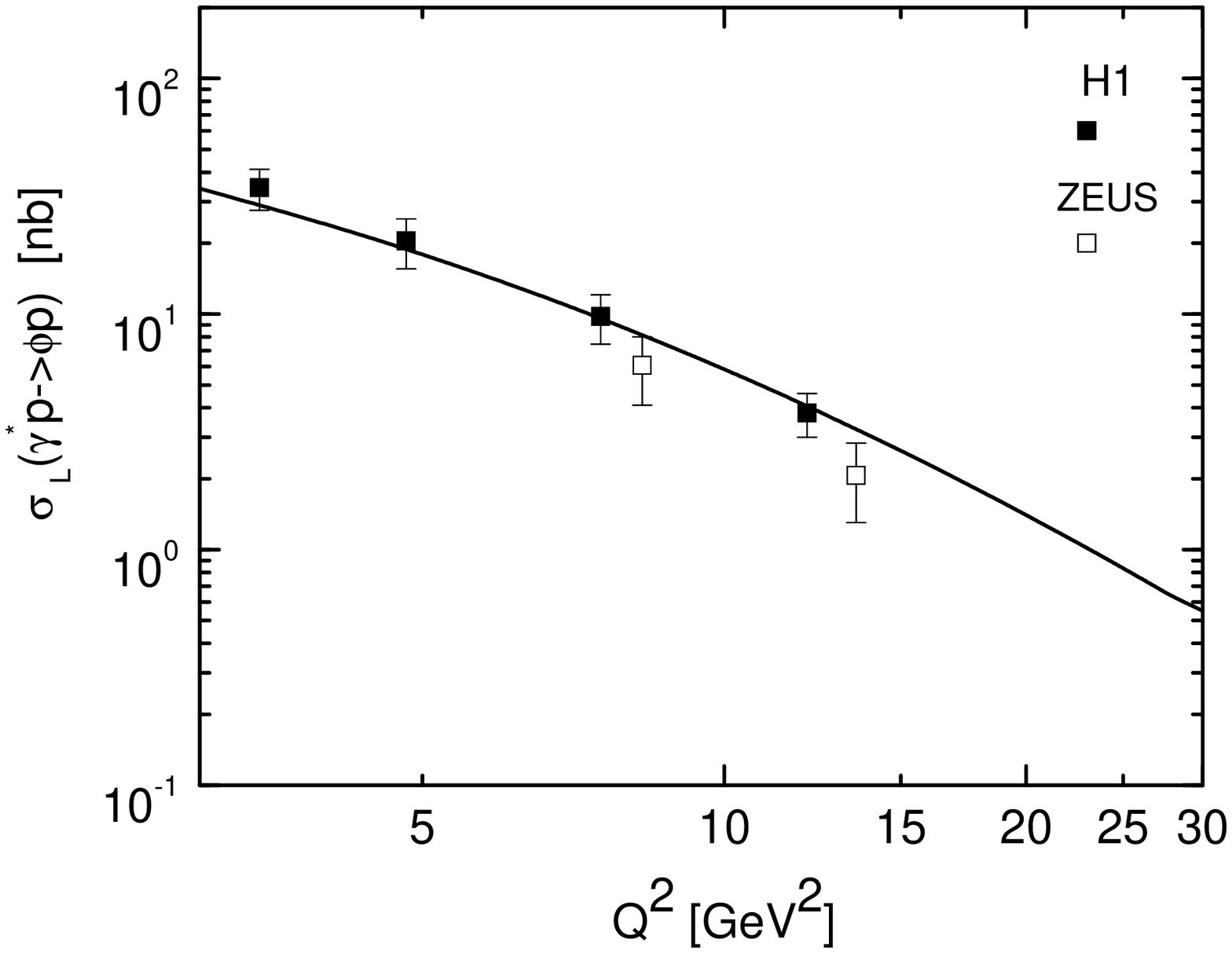}\hspace*{0.3cm}
\includegraphics[width=0.44\textwidth, bb= 17 348 532 741,clip=true]
{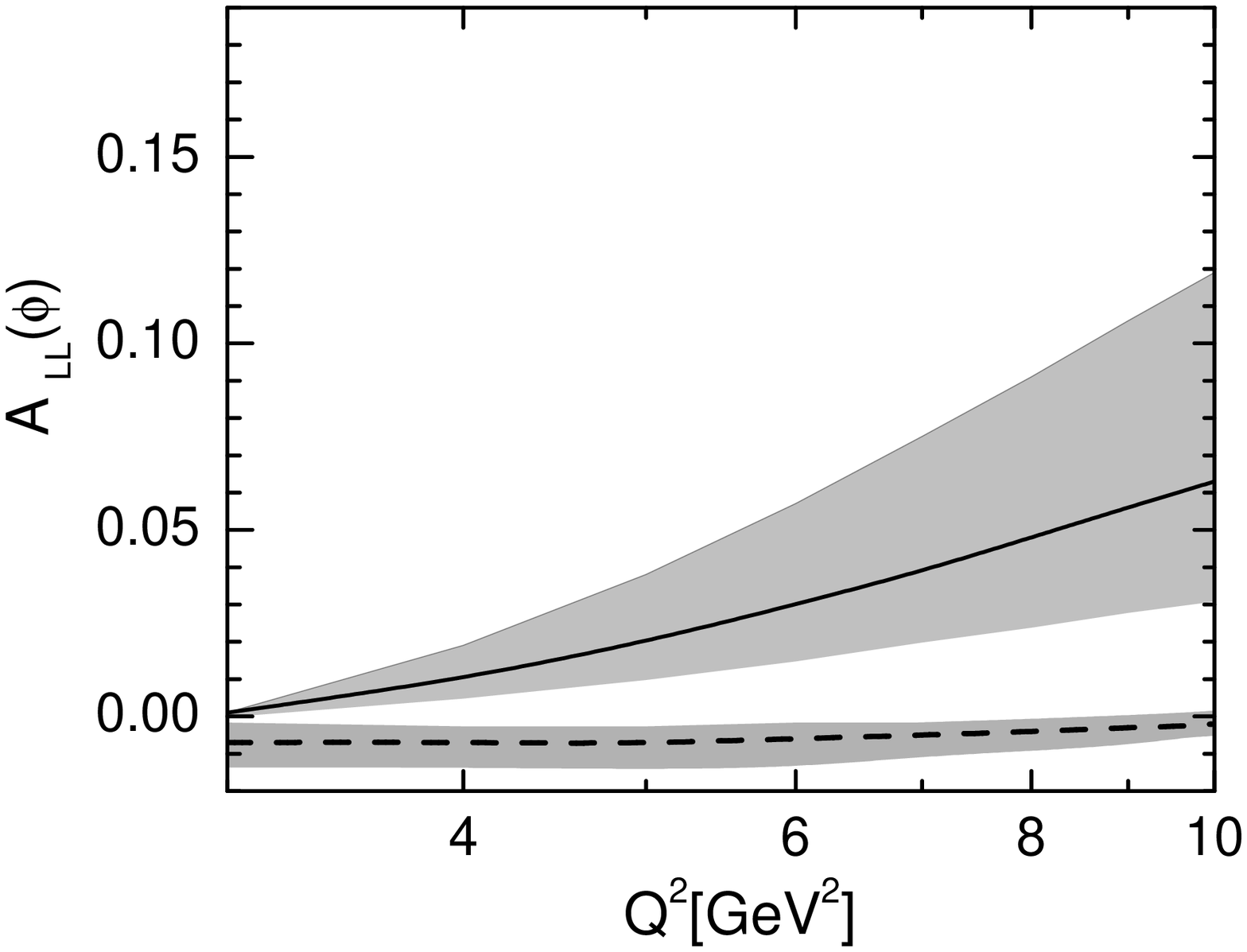}
\caption{Left: The integrated cross section for $\gamma_L p\to \phi p$
  versus $Q^2$ at $W\simeq 75\, \gev$. Data taken from \protect\ci{h1}
  (filled squares) and \protect\ci{zeus96} (open symbols). The solid
  line represents the result obtained in Ref.~\protect\ci{golo}.
  Right: Predictions for the helicity correlation $A_{LL}$ for $\phi$ 
  electroproduction versus $Q^2$ at $W=5\,\gev$ (solid line) and
  $W=10\,\gev$ (dashed line), $t\simeq 0$ and $y\simeq 0.6$. The
  shaded bands reflect the uncertainties due to the errors of the
  gluon distributions.}
\label{fig:4}
\end{figure}

The handbag approach leads to the following proton helicity
non-flip amplitude
\be
M^\phi_{\mu' +,\mu +}\= -\frac{e}{6}\, \int_0^1\,
                            \frac{d\xb}{(\xb+\xi)(\xb-\xi+i\varepsilon)}\,
                 \Big[H^\phi_{\mu' +,\mu +} + H^\phi_{\mu' -,\mu -}\Big]\,
                 H^g(\xb,\xi,t)\,.
\label{ampl}
\ee
Contributions from other GPDs can be neglected at small $\xbj$ and for
unpolarized protons. The photon and meson helicities are denoted by
$\mu$ and $\mu'$, respectively. The explicit labels in the full 
(subprocess) amplitude, $M^\phi$ ($H^\phi$), refer to the helicities of
the protons (gluons).\\
The GPDs are controlled by non-perturbative QCD. In the absence of a GPD 
analysis in analogy to those of the usual PDFs (see however Ref.~\cite{DFJK4}) 
one has to rely on a model. Following other authors we restrict
ourselves to the forward direction and exploit the ansatz for a double 
distribution proposed in Ref.\ \ci{musa} ($n=1,2$)
\be
f(\beta,\alpha,t\simeq 0) \= g(\beta)\,
                 \frac{\Gamma(2n+2)}{2^{2n+1}\,\Gamma^2(n+1)}\,
               \frac{[(1-|\beta|)^2-\alpha^2]^n}{(1-|\beta|)^{2n+1}} \,.
\label{double-dis}
\ee 
The GPD is then obtained by an integral over $f$
\be
H^{g}(\xb,\xi) = \Big[\,\Theta(0\leq \xb\leq \xi)
         \int_{x_3}^{x_1}\, d\beta + 
       \Theta(\xi\le \xb\leq 1) \int_{x_2}^{x_1}\, d\beta \,\Big] 
        \frac{\beta}{\xi}\,f(\beta,\alpha=\frac{\xb-\beta}{\xi})\,.
\label{gpd-model}
\ee
Using the NLO CTEQ5M \ci{CTEQ} results on the gluon distribution as 
input one obtains the GPD $H^g$ shown in Fig.\ \ref{fig:1}. While 
the double-distribution ansatz guarantees polynomiality of the moments 
of $H$, positivity is not automatically satisfied. It can however 
be checked that the positivity bounds are respected by the model
\req{double-dis}, \req{gpd-model} 
in the $\xi$ and $\xb$ range of interest. Inspired by the Regge model 
one may generalize to non-zero but small $t$ by a factor 
$\exp{[(\alpha' \ln(1/\beta)+b_0)t]}$ in \req{double-dis} where 
$\alpha'$ is the slope of the Pomeron trajectory. The investigation 
of this $t$ dependence is left to a forthcoming publication.\\
The last item of the amplitude \req{ampl} to be discussed is the
subprocess amplitude. Its treatment is rather standard, it only
differs in detail from versions to be found in the literature
\ci{martin,frank}. In the modified perturbative approach invented by
Sterman and collaborators \ci{li92}, in which quark transverse momenta
are retained and gluonic radiative corrections in the form of a
Sudakov factor are taken into account, it reads
\be
{\cal H}^\phi\=\int \frac{d\tau d k^2_\perp}{\sqrt{2}16\pi^2} \Psi_\phi 
    \,{\rm Tr} \left\{(q\sla'+m_\phi)\eps_\phi\sla^*T_0 
       - \frac{k^2_\perp g_\perp^{\alpha\beta}}{2M_\phi}\, 
       \{(q\sla'+m_\phi)\eps_\phi\sla^*,\gamma_\alpha\}\, 
            \Delta T_\beta\right\}\,.
\label{amp-mpa}
\ee
Higher order terms in this expansion are not shown. Gaussians for the 
meson wavefunctions, $\Psi_\phi=\Psi_\phi(\tau,k^2_\perp)$, are used
which may depend on the polarization of the vector meson. The first
term in \req{amp-mpa} dominates for $\phi_L$ while it is approximately 
zero for transversally polarized vector mesons ($\phi_T$). The second 
term dominates in this case. The soft physics parameter $M_\phi$ in
the second term of Eq.~\req{amp-mpa} is of order of the vector meson 
mass $m_\phi$. As can be seen from Eq.~\req{amp-mpa} the $L\to L$ 
transition is dominant while the $T\to T$ one is of relative order
$\langle k_\perp^2\rangle^{1/2}/Q$ and the $T\to L$ one of order
$\sqrt{-t}/Q$. The latter amplitude is tiny and only noticeable in
some of the SDMEs. All other transitions are negligible.\\
Before comparing the results to experiment a comment is in order on the
$t$ dependence of the amplitudes. Exponentials in $t$ are assumed with 
slopes $B^\phi_{LL(TT)}$ taken from experiment. In combination with the 
calculated forward amplitudes one can thus evaluate the integrated 
cross sections and the SDME for small $t$. From \req{amp-mpa} one sees
that the size of the $T\to T$ amplitude is controlled by the following 
product of parameters ($f_T^\phi$ denotes the corresponding decay constant) 
\be
\Big| M_{TT}^\phi\Big| \propto \left(\frac{f_T^\phi}{M_\phi}\right)^2\,
\frac1{B^\phi_{TT}}\,.
\ee 
Since the available data do practically not allow for an independent 
determination of the slope $B^\phi_{TT}$, only this product is probed. 
Only the SDMEs are slightly sensitive to the value 
of $B_{TT}^\phi$. All values of this slope lying in the range from about
$B_{LL}^\phi/2$ up to about $B_{LL}^\phi$ lead to fair agreement with
the present data~\cite{h1,zeus96,zeus05}.\\
In Fig.~\ref{fig:3} a selection of SDMEs is shown and the results
obtained in Ref.~\ci{golo} are compared to experiment~\ci{h1,zeus05}.  
Results for the integrated cross section $\sigma_L$ are displayed  
and compared to data in Fig.~\ref{fig:4}.
The GPD approach as detailed in Ref.~\ci{golo}, can in principle also
be applied to electroproduction of $\phi$ mesons for COMPASS kinematics 
where $W$ is much smaller than at HERA. Contributions from the quark
GPDs are expected to be very small due to the mismatch of the $\phi$
and proton valence quarks. Even for HERMES kinematics the quark
contribution is likely tiny. As an example the initial state helicity 
correlation $A_{LL}$ is shown in Fig.~\ref{fig:4} which measures an 
interference term between the contribution from the GPD $H^g$ (see 
Eq.~\req{amp-mpa}) and a similar one from the GPD $\widetilde{H}^g$. 
The latter contribution is negligible in the cross sections and SDMEs, 
the relative size of $\widetilde{H}^g$ and $H^g$ is approximately 
given by the ratio of the polarized and unpolarized gluon
distributions. The smallness of this ratio leads to small values of
the helicity correlation. A detailed investigation of $\phi$ 
electroproduction in the GPD framework for COMPASS and HERMES
kinematics is in progress.\\
The approch discussed here applies also to $\rho$ electroproduction for
HERA kinematics. Indeed results for this process are given in
Ref.~\ci{golo}. The analysis of electroproduction of $\rho$ mesons at 
COMPASS and HERMES kinematics definitely requires the inclusion of the 
quark contributions. 

I summarize: $\phi$ meson electroproduction off unpolarized protons
at small $\xbj$ and small $t$ probes the GPD $H^g$. Calculating the
partonic subprocess within the modified perturbative approach (using
gaussian wavefunctions), one achieves fair agreement with HERA data on the
integrated cross sections for longitudinally and transversally
polarized virtual photons and the SDMEs for electroproduction of
$\phi$ mesons. It is to be stressed that only the forward amplitudes 
are caluclated within the GPD approach as yet. 
Their $t$ dependencies are assumed to be exponentials with slopes
taken from experiment. The present data do, however, not fix the
slope of the $T\to T$ amplitude precisely. This treatment of the $t$ 
dependence is unsatisfactory and improvements are required. In
principle the GPD approach has the potential to do better but the GPDs 
as a function of $t$ are needed for that. Some results on $\phi$ 
production for COMPASS kinematics are already presented in Ref.~\ci{golo}.\\
Acknowledgments: This work has been partially supported by the
Integrated Infrastructure Initiative 'Hadron Physics' of the European
Union, contract No. 506078.

\end{document}